\documentclass[two column, prA]{revtex4}%
\usepackage{amsmath}
\usepackage{graphicx}
\usepackage{amsfonts}
\usepackage{amssymb}
\usepackage{color}%
\providecommand{\U}[1]{\protect\rule{.1in}{.1in}}
\ifx\pdfoutput\relax\let\pdfoutput=\undefined\fi
\newcount\msipdfoutput
\ifx\pdfoutput\undefined\else
\ifcase\pdfoutput\else
\ifx\paperwidth\undefined\else
\ifdim\paperheight=0pt\relax\else\pdfpageheight\paperheight\fi
\ifdim\paperwidth=0pt\relax\else\pdfpagewidth\paperwidth\fi
\fi\fi\fi
\begin{document}
\title{Reply to arXiv:2203.14555}
\author{Margaret Hawton}
\affiliation{Department of Physics, Lakehead University, Thunder Bay, ON, Canada, P7B 5E1}

\begin{abstract}
This comment refutes the claim made by A. Jadczyk and A.M. Schlichtinger in
arXiv:2203.14555 that the photon position operator with commuting components
proposed in 1999 does not have the properties required for a photon position operator.

\end{abstract}
\maketitle

All of the statements in \cite{JadezykSchliechtinher} relating to photon
position operators with commuting components (PPOCC) are incorrect since they
are based on the assumption of spherically symmetry that is appropriate for
massive particles but not for the photon. The symmetry of a particle's states
is determined by Wigner's little group that can be boosted to give all
acceptable states. For mass positive definite particles this little group is
the three-dimensional rotation group, while for the zero mass photons it is a
single rotation that can be called $\widehat{J}_{3}$ and two generators of
gauge transformations. This is discussed in \cite{Weinberg} and
\cite{WignerLittleGroup}. The mathematical proofs referred to in paragraphs 1
and 2 of the Introduction to \cite{JadezykSchliechtinher} and the properties
referred to in its Abstract are based on the assumption of spherical symmetry
and hence do not apply to the photon.

There is no mathematical or logical error in \cite{WignerLittleGroup},
\cite{HawtonBaylis} or \cite{HawtonPosOp}. The contents of these refereed and
published papers is completely consistent with the Poincar\'{e} algebra
$\left[  \widehat{J}_{i},\widehat{J}_{j}\right]  =i\hbar\epsilon
_{ijk}\widehat{J}_{k},$ $\left[  \widehat{J}_{i},\widehat{K}_{j}\right]
=i\epsilon_{ijk}\widehat{K}_{k},$ $\left[  \widehat{K}_{i},\widehat{K}%
_{j}\right]  =-i\hbar\epsilon_{ijk}\widehat{J}_{k},$ $\left[  \widehat{J}%
_{i},\widehat{P}_{j}\right]  =i\hbar\epsilon_{ijk}\widehat{P}_{k},$ $\left[
\widehat{K}_{i},\widehat{P}_{j}\right]  =i\hbar\delta_{ij}\widehat{H},$
$\left[  \widehat{K}_{i},\widehat{H}\right]  =-i\hbar\widehat{P}_{i},$
$\widehat{J}_{i},\widehat{H}=\left[  \widehat{P}_{i},\widehat{H}\right]
=\left[  \widehat{P}_{i},\widehat{P}_{j}\right]  =0$ for $i=1,2,3$ satisfied
by the conserved total momentum, total angular momentum and boost operators.
There is no separation into spin and orbital angular momentum and no position
operator in this general Poincar\'{e} algebra because these physical
quantities are not conserved. In \cite{WignerLittleGroup} and
\cite{HawtonBaylis} a realization of the Lie algebra of the Poincar\'{e}
group" is specified in which total angular momentum is separated into its
extrinsic and intrinsic parts as $\widehat{\mathbf{J}}=\widehat{\mathbf{x}%
}^{\left(  \chi\right)  }\times\hbar\mathbf{k}+\widehat{\mathbf{J}}^{\left(
0,\chi\right)  }$ where $\widehat{\mathbf{x}}^{\left(  \chi\right)  }$ is a
PPOCC and $\widehat{\mathbf{J}}^{\left(  0,\chi\right)  }$ is the intrinsic
angular momentum.

Prior to the work described above it was assumed that photon position
eigenvectors, like those describing massive particles, are spherically
symmetric. As discussed in the first paragraph this assumption is not
appropriate for photons. Ref. \cite{HawtonPosOp} is just a contruction of the
PPOCC that proves its existence. The symmetry of its eigenvectors and its
implementation of the Poincar\'{e} algebra is discussed in \cite{HawtonBaylis}
and its relationship to the Wigner little group is established in
\cite{WignerLittleGroup}.

\end{document}